\documentclass[twocolumn,aps,prb,superscriptaddress]{revtex4-1}

\usepackage{amsmath,amssymb,graphicx}
\usepackage{soul}
\usepackage{caption}
\usepackage{graphicx}
\usepackage{color}
\usepackage{ulem}

\usepackage{sidecap}
\usepackage{floatflt}

\begin{document}

\bibliographystyle{naturemag}

\title{Controlling the optical spin Hall effect with light}


\author{O. Lafont}
\affiliation{Laboratoire Pierre Aigrain, \'Ecole Normale Sup\'{e}rieure, PSL Research University,
CNRS, Universit\'{e} Pierre et Marie Curie, Sorbonne Universit\'es, Universit\'{e} Paris Diderot,
 Sorbonne Paris Cit\'{e},
24 rue Lhomond, 75231 Paris Cedex 05, France}

\author{M.H. Luk}
\affiliation{Department of Physics, University of Arizona, Tucson, AZ 85721 USA}

\author{P. Lewandowski}
\affiliation{Physics Department and Center for Optoelectronics and Photonics Paderborn (CeOPP), Universit\"at Paderborn, Warburger Strasse 100, 33098 Paderborn, Germany}

\author{N.H. Kwong}
\affiliation{College of Optical Sciences, University of Arizona, Tucson, AZ 85721 USA}

\author{K.P. Chan}
\affiliation{Department of Physics and Institute of Theoretical Physics, The Chinese University of Hong Kong, Hong Kong SAR, China}

\author{M. Babilon}
\affiliation{Physics Department and Center for Optoelectronics and Photonics Paderborn (CeOPP), Universit\"at Paderborn, Warburger Strasse 100, 33098 Paderborn, Germany}

\author{P.T. Leung}
\affiliation{Department of Physics and Institute of Theoretical Physics, The Chinese University of Hong Kong, Hong Kong SAR, China}

\author{E. Galopin}
\affiliation{Laboratoire de Photonique et de Nanostructures, CNRS Route de Nozay, FR-91460 Marcoussis, France}

\author{A. Lemaitre}
\affiliation{Laboratoire de Photonique et de Nanostructures, CNRS Route de Nozay, FR-91460 Marcoussis, France}

\author{J. Tignon}
\affiliation{Laboratoire Pierre Aigrain, \'Ecole Normale Sup\'{e}rieure, PSL Research University,
CNRS, Universit\'{e} Pierre et Marie Curie, Sorbonne Universit\'es, Universit\'{e} Paris Diderot,
 Sorbonne Paris Cit\'{e},
24 rue Lhomond, 75231 Paris Cedex 05, France}

\author{S. Schumacher}
\affiliation{Physics Department and Center for Optoelectronics and Photonics Paderborn (CeOPP), Universit\"at Paderborn, Warburger Strasse 100, 33098 Paderborn, Germany}

\author{E. Baudin}
\affiliation{Laboratoire Pierre Aigrain, \'Ecole Normale Sup\'{e}rieure, PSL Research University,
CNRS, Universit\'{e} Pierre et Marie Curie, Sorbonne Universit\'es, Universit\'{e} Paris Diderot,
 Sorbonne Paris Cit\'{e},
24 rue Lhomond, 75231 Paris Cedex 05, France}

\author{R. Binder}
\affiliation{College of Optical Sciences, University of Arizona, Tucson, AZ 85721 USA}
\affiliation{Department of Physics, University of Arizona, Tucson, AZ 85721 USA}




\begin{abstract}
The optical spin Hall effect (OSHE)
is  a transport phenomenon of exciton polaritons in semiconductor microcavities,
caused by the polaritonic spin-orbit interaction,
that leads to the formation of spin textures.
In the semiconductor cavity, the physical basis of the spin orbit coupling
 is an effective magnetic field caused by the splitting of transverse-electric and transverse-magnetic (TE-TM) modes. The spin textures
 can be observed in the near field (local spin distribution of polaritons), and as light polarization patterns in the more readily observable far field.
 For future applications in spinoptronic devices,  a simple and robust control mechanism, which establishes a one-to-one correspondence between stationary incident light intensity and far-field polarization pattern, is needed. We present such a control scheme, which is made possible by a specific double-microcavity design.

 \end{abstract}

\maketitle


The detection and manipulation of spins is an important part of science, in areas ranging from quantum computing, information, and spintronics\cite{awschalom-etal.02}, to ubiquitous medical imaging techniques such as Magnetic Resonance Imaging (MRI).
Much of the functionality of spin effects rests on the ability to control the spin dynamics through the application of external optical and/or magnetic fields. For example, in MRI spatial gradients of external magnetic fields control the
spin precession, and in electron or nuclear spin-based quantum computing logical operations are performed
using spatially or time-varying electromagnetic fields
 (e.g.\ Ref.\ \cite{kane.98,awschalom-etal.02}).
Major research efforts have focused on photonic counterparts to magnetic spin systems, including
the plasmonic spin Hall effect \cite{xiao-etal.15}, and, importantly, wide-ranging investigations of the
spin orbit interactions of light
\cite{%
liberman-zeldovich.92,%
brasselet-etal.09,%
rodriguez-herrera-etal.10,%
yin-etal.13,%
bliokh-etal.15%
}.
All-optical spin systems  combine the benefits of magnetic spin
systems and their (sometimes relatively simple) spin dynamics with the highly developed technology of optical preparation and detection
of polarization states (the optical analogue of spin).

 A promising semiconductor system is a microcavity containing semiconductor quantum wells, where spin states of exciton-polaritons can be created optically and the  TE-TM splitting yields a spin-orbit interaction that can be described by an effective magnetic field.
  This,  in turn,  gives rise to a polaritonic spin Hall effect,
 the so-called optical spin Hall effect (OSHE)
 \cite{kavokin-etal.05,%
leyder-etal.07,%
langbein-etal.07,%
manni-etal.11,%
maragkou-etal.11,%
kammann-etal.12}.
Since polaritons with different in-plane wave vectors
experience different effective magnetic fields,
an  isotropic distribution of polaritons on a ring in wave vector space
can lead to an anisotropic polarization texture or pattern, both in real (configuration) and wave vector space.
Such polarization/spin textures
have been found for excitations of linearly  and circularly  polarized polaritons in Ref.\
\cite{leyder-etal.07} and \cite{kammann-etal.12}, respectively
  (structurally similar polarization/spin textures are also present in different physical systems, e.g.\
\cite{hielscher-etal.97,schwartz-dogariu.08}).
The OSHE in wave vector space  corresponds to that seen experimentally in far-field observations, which are particularly important in possible photonics or spinoptronics\cite{shelykh-etal.04,shelykh-etal.10} applications. In the following, we
demonstrate the all-optical manipulation and well-defined control of the polarization/spin texture. Our set-up is conceptually simple, allowing for robust steady state control that would be required in future device applications.
 We show that a circularly polarized pump rotates the far-field polarization/spin texture.
  The rotation angle generally increases with the pump spot size, and for Gaussian beam profiles its dependence on the pump power
 approaches a simple arctangent law in the limit of large spot sizes. \\

{\centerline {\bf Results}}

\begin{figure}
\vspace*{5cm}
\includegraphics[scale=0.6]{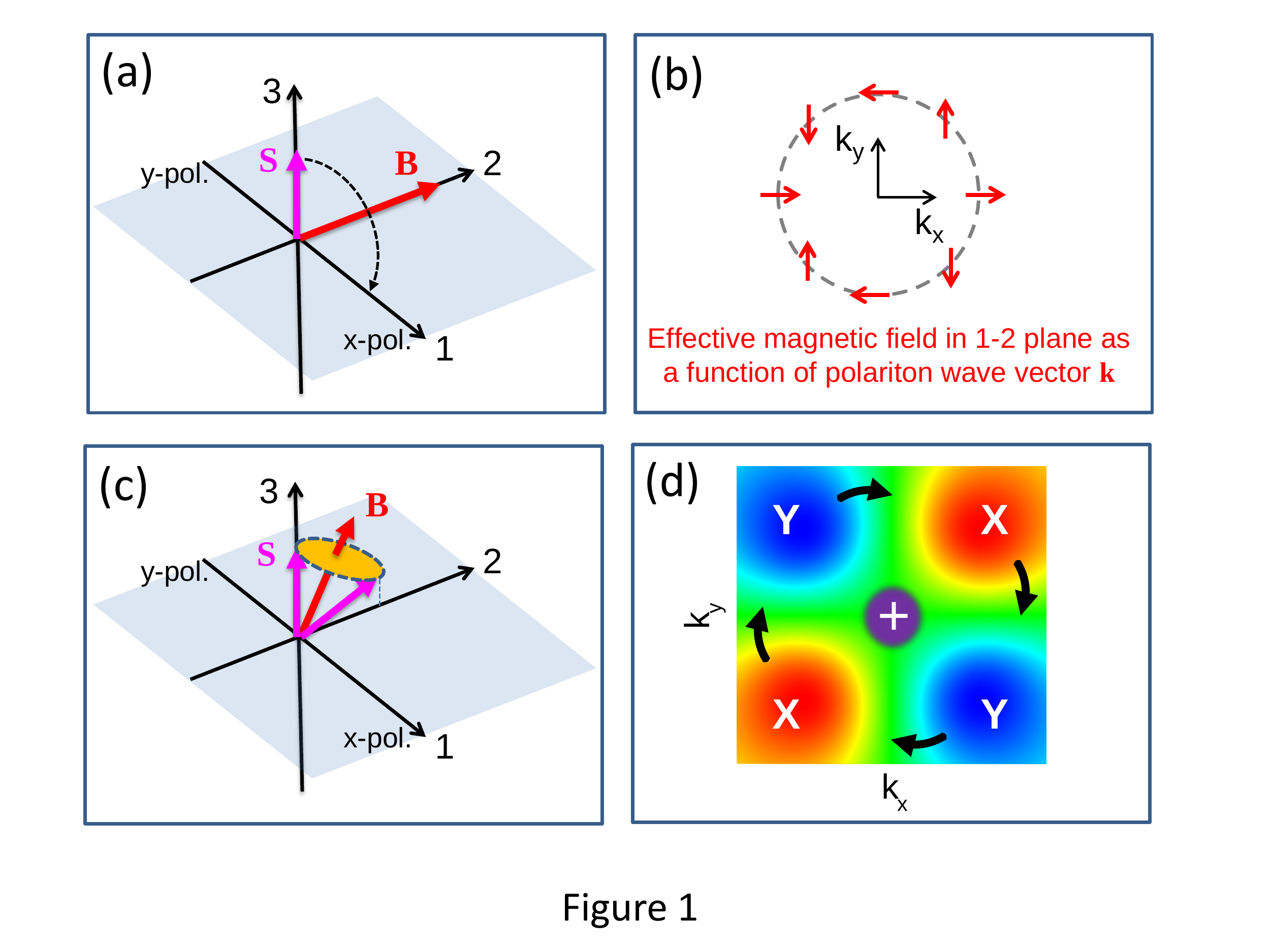}
\caption{
{\bf Schematics of the optical spin Hall effect.}
(a) Sketch of pseudo-spin space and torque action when $\textbf{B}$ has only a positive $B_2$ component (corresponding to polariton wave vector $\textbf{k}$ along the $45^{\circ}$ direction (see (b)).
(b) Sketch of the effective magnetic field vectors (red) for different directions of the polariton wave vector $\textbf{k}$.
(c) Similar to (a), but for a $\textbf{B}$ with a non-zero $B_3$ component.
(d) Sketch of the steady state $S_1$ component of the pseudo-spin vector as a function of the polariton wave vector $\textbf{k}=(k_x,k_y)$. Red (blue) color corresponds to x (y) polarization. The circularly (``+'') polarized pump at $\textbf{k}=0$ (normal incidence) is also indicated. The black arrows indicate the rotation of the pattern that occurs when $B_3$ is non-zero and positive.
}
\label{fig:steady-state-torque}
\end{figure}

Our theoretical explanation of the experiment is based on solving microscopic equations of motion of the polaritons. However, to gain insight into the physical nature of the optical spin Hall effect, it is helpful to begin the discussion by
considering a simple pseudo-spin model\cite{kavokin-etal.05,leyder-etal.07}. The pseudo-spin vector $\textbf{S}$ is given in terms of
the Stokes parameters formed by the polariton wave functions, and its precise definition becomes important and will be given below. For our initial discussion it is sufficient to note that the pseudo-spin model highlights the effect of the spin-orbit interaction on quantities that are
 readily observable in the far field, namely the components of $\textbf{S}(\textbf{k})$, where $\textbf{k}$ is the polariton's in-plane wave vector.
The evolution of $\textbf{S}$ is given by the following torque equation,
%
%
\begin{equation}
 \dot {\textbf{S} } (\textbf{k}) = \textbf{B}(\textbf{k} ) \times \textbf{S}(\textbf{k})
 - \gamma \textbf{S}(\textbf{k})    +  \textbf{R}(\textbf{k})
 \end{equation}
 which includes a decay term (with decay rate $\gamma$) and a source  $\textbf{R}$.
 %
 As will become clear from the definition of $\textbf{S}$  given below,
 positive (negative) $S_1$ corresponds to dominant x (y) linear polarization components of the polariton fields, and positive (negative) $S_3$ to ``+'' (``-'') circular polarizations. Throughout this paper, we assume ``+'' circularly polarized pump, and hence
 $\textbf{R} = (0, 0, R)$.
 In steady state, we find
 $S_1 = A (\gamma B_2 + B_1 B_3)$ with
 $A=R(\textbf{B}^2 + \gamma^1)^{-1}$.
 In the low density limit, the magnetic field
 is restricted to the
 1-2 plane of the three-dimensional pseudo-spin space ($B_3=0$), and has the angle dependence
$\textbf{B}(\textbf{k} ) = \Delta_{k} (\cos(2 \phi ), \sin(2 \phi), 0)$, where $\Delta_{k}$ is the TE-TM splitting and $\phi$ the polar angle of the vector $\textbf{k}$. The example of $\phi = 45^{\circ}$ is sketched in
 Fig.\ \ref{fig:steady-state-torque}a:
 the torque of the B-field having only a positive $B_2$ component (see
Fig.\ \ref{fig:steady-state-torque}b)
rotates $\textbf{S}$ towards positive $S_1$, i.e.\ toward x-polarization until it settles to steady state (in this case in the 1-3 plane).
At $\phi = -45^{\circ}$, the negative $B_2$ component of $\textbf{B}$
 would favor y-polarization. Indeed, for $B_3 =0$ we have
 $S_1 = A \gamma B_2$, creating a simple and by now well-known map between the angular dependence of $B_2$ and that of $S_1$.

For future applications of the OSHE a direct optical control scheme is desirable. Manipulations of the OSHE patterns have been discussed theoretically \cite{flayac-etal.13prl}, and experimental manipulations and transient effects in near-field patterns have recently been reported in Ref,\ \cite{cilibrizzi-etal.15}.
  Our present work lays out a path toward a robust approach to an all-optical control of the stationary OSHE pattern. The optical excitation scheme needs to be simple in the sense that the spatial distribution of optically-excited polaritons coincides with the spatial optical pump beam profile and that there is no need for a polariton condensate. Importantly, our goal is to
 have a simple and robust one-to-one correspondence between optical pump power and the modification of the far-field polarization pattern.
As we show below, we meet these objectives by using a cavity that provides two polariton branches, one that contains the
 off-axis ($\textbf{k} \neq 0$) polaritons, which actually
form the OSHE pattern and which do not couple directly to the external field, and
 one that contains the $\textbf{k} = 0$ polaritons  that are directly pumped by an external field in
normal incidence. Choosing a circularly polarized optical pump, we are then able to create substantial densities of circularly polarized polaritons. As a result of Coulombic polariton interactions, the optical excited polaritons
create a $B_3$ component of the effective magnetic field. This has been found in Refs.\ \cite{lagoudakis-etal.02,shelykh-etal.10} for the case of a single cavity. Below, we will generalize the interactions to the case of a double cavity, which is essential for  our excitation scheme. For the present simplified pseudo-spin argument, it suffices to assume  $B_3 \neq 0$ as a consequence of the optical pumping. Then the nutation of an initially created pseudo-spin along the 3-direction is more complex, as schematically shown in
Fig.\  \ref{fig:steady-state-torque}c.
Now the steady state solution is
$S_1 = A (\gamma B_2 + B_1 B_3) \sim \cos (2 \phi + \Theta)$
where
$\tan \Theta = - \gamma / B_3$. This shows that increasing $B_3$, i.e.\ increasing the optical pumping, rotates the stationary pseudo-spin texture by the angle $\Theta$, as indicated by the black arrows in
Fig.\  \ref{fig:steady-state-torque}d,
where we use the source as a $\textbf{k}$-independent parameter.
Below,  we use a full microscopic theory to corroborate the basic predictions of the simple pseudo-spin model, identify $B_3$ in terms of the polariton wave functions, and clarify the properties of the source $R$, which creates polaritons on a ring (the elastic circle) in $\textbf{k}$ space.

Before presenting more details of the microscopic theory, let us briefly summarize key aspects of semiconductor microcavities.
In their simplest form, such cavities consist of one (or several) semiconductor quantum wells between two mirrors (such as distributed Bragg reflectors, DBR). Excitons, which are optically excited quasi-particles in the quantum well, interact with the light field inside the cavity to form exciton polaritons. These
 polaritons, which have distinct advantages such as long decoherence times,  form quantum fluids with spinor field characteristics.



The excitonic component of the polariton provides a strong Coulombic spin-dependent exciton-exciton interaction which is conceptually similar to the Heitler-London model for chemical bonds in hydrogen molecules,
where the sum (difference) of direct and exchange interactions determine the singlet (triplet) molecular energies.
For excitons, this results in interactions $T^{++}$ ($T^{+-}$) that are different if the two excitons have same (opposite) circular polarization.
For quantum well excitons, these interactions have been
 quantitatively evaluated in Ref.\
 \cite{takayama-etal.02}, and  for polaritons they result in a spin and energy-dependent interaction\cite{schumacher-etal.07prb}. Previously, these interactions have been found to be important in  spin-dependent optical nonlinearities (e.g.\ Ref.\ \cite{kwong-etal.01prl,lecomte2014}),
Bose-Einstein condensates (for a review, see e.g.\ Ref.\ \cite{deng-etal.10}),
spin-dependent optoelectronic (sometimes referred to as spinoptronic) devices (e.g.\ Ref.\ \cite{shelykh-etal.04,shelykh-etal.10}), and polaritonic Feshbach resonances\cite{takemura-etal.14}.
In the present context, they modify the effective spin-orbit interaction by modifying the effective magnetic field\cite{shelykh-etal.10}, and hence the spin and polarization texture of the polariton field in real and wave vector space\cite{flayac-etal.13prl}.

\begin{figure}
\includegraphics[scale=0.6]{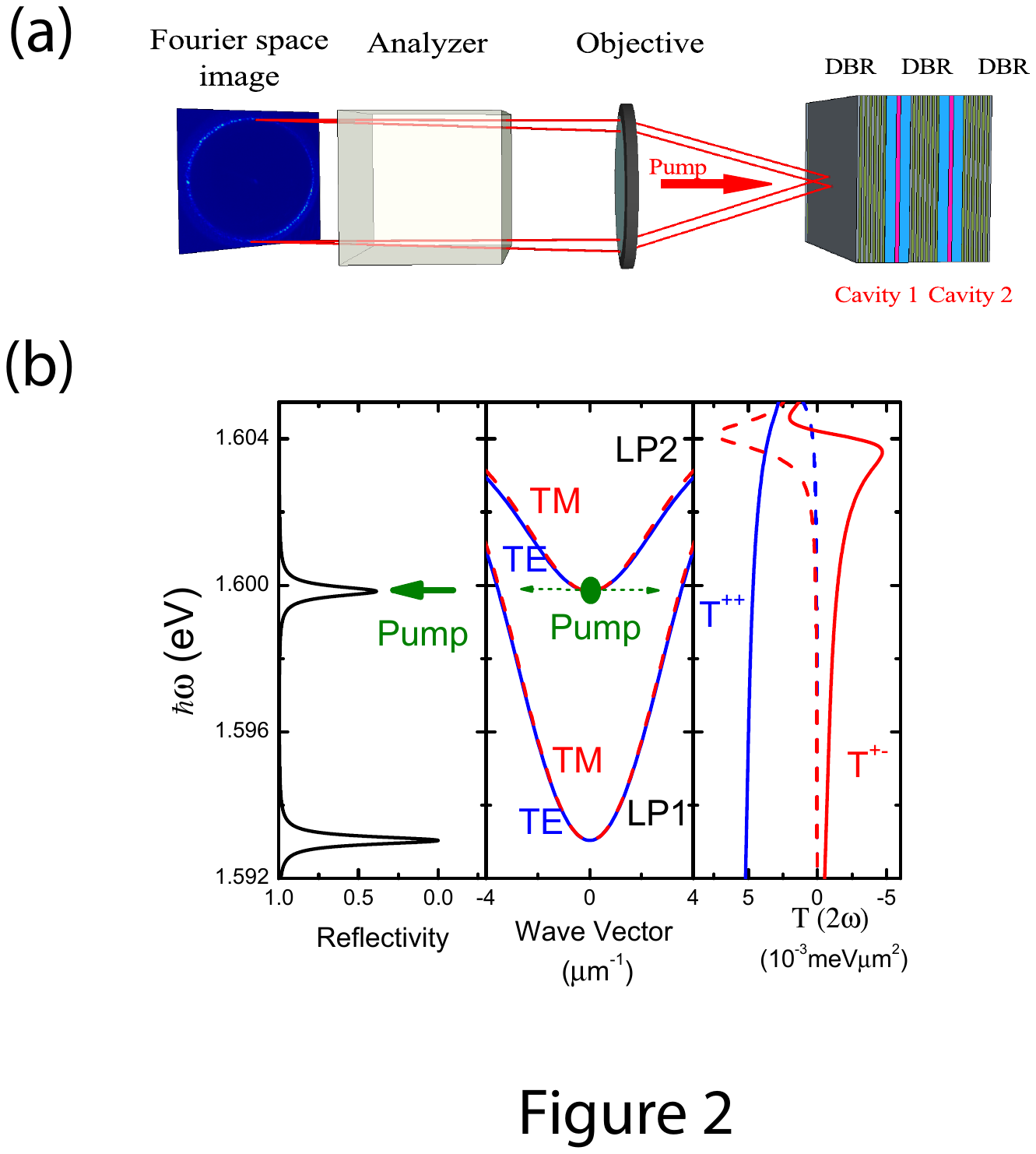}
\caption{
{\bf The double cavity.}
(a) Sketch of the experimental setup with double cavity, normal-incidence pump and detection of the polarization map in reflection geometry.
(b) The frequency-dependent reflectivity, dispersion relations of the two lower polariton branches, and exciton-exciton interactions
 $T^{++}$ and $T^{+-}$ from Ref.\ \cite{takayama-etal.02}. The branch LP2 allows the normal incidence pump to enter the cavity. The LP2 polaritons act as a source for LP1 polaritons, as indicated by the dashed horizontal arrows, and simultaneously as a polariton reservoir that controls the orientation of the OSHE pattern.
 The  splitting between TE and TM polaritons, particularly large in our double-cavity structure,
  results in spin-orbit coupling via an effective magnetic field\cite{kavokin-etal.05}.
  Modification of the effective magnetic field is enabled through optical pumping and the resulting spin-dependent interaction between polaritons, which in turn is based on the underlying exciton-exciton interaction.
}
\label{fig:cavity-dispersion}
\end{figure}

 As pointed out above, a circularly (``+'') polarized pump beam results in the tilting of the effective magnetic field $\textbf{B}(\textbf{k})$  out of the 1-2 plane in pseudo-spin space, and hence can lead to a modification of the spatial polarization texture of the polaritons' linear polarization  components, Fig.\
 \ref{fig:steady-state-torque}.
%
%
 However, a simple implementation of such an optical pump scheme,
 which involves only a single spatial component and a quasi-monochromatic pump acting both as the source of the polaritons involved in the OSHE and their external control,
 is difficult in conventional single-cavity designs. A double-cavity design containing two lower polariton branches can overcome those difficulties.
The reflectivity spectrum of the structure we use is shown in
Fig.\
\ref{fig:cavity-dispersion}.
 Owing to the two coupled cavities, this system features two lower polariton branches,
  Fig.\ \ref{fig:cavity-dispersion}b,
  with the upper of the two, LP2, yielding a reflectivity window that allows a normal-incidence monochromatic pump to enter the cavity, creating LP2 polaritons. Secondly,
 it creates a population of LP1 polaritons (mostly via Rayleigh scattering, but, depending on the pump beam profile, possibly also through overlap of the pump's spatial frequencies with the LP1 dispersion).
%
 %
 Due to the underlying exciton-exciton interactions between LP1 polaritons on the elastic circle and LP2 polaritons close to $k=0$,
 shown in
 Fig.\ \ref{fig:cavity-dispersion}b,
 the pump can induce a shift of the LP1 frequencies.
 In our case, the shift is different for ``+'' and ``-'' polarized LP1 states, which is important for the control of the OSHE.
 We note that the shifts are substantial (of order 0.5meV)
  even though in the present study we stay
  below the optical parametric oscillation (OPO) threshold.
%
We note again that,
in contrast to previous works,
(e.g.\ Ref.\ \cite{kammann-etal.12,cilibrizzi-etal.15}),
in our approach the modification of the effective magnetic field, which is determined by exciton-exciton interactions,
 does not require other nonlinear effects such  as the presence of a Bose condensate.
Finally, an additional benefit of our double-cavity is its
particularly large TE-TM splitting of about 0.2 meV at our pump frequency (compared to about 0.05 meV in Ref.\ \cite{leyder-etal.07}).

In addition to exciton-exciton interactions, TE-TM splitting provides a coupling mechanism for the
``+'' and ``-''  states, i.e.\ a spin-orbit interaction.
  It leads to an anisotropic wave interference that, in turn, produces a polarization texture of the polaritonic quantum fluid inside the cavities (OSHE). In the far field (here measured in reflection) such a polarization texture can be observed through polarization sensitive angle-resolved detection. Before discussing our experimental results for the far-field polarization map and its modification at higher pump fluences, we lay out the theoretical predictions.

Spinor-valued polaritons are represented by wave functions $\psi^{\pm}_m(\textbf{r},t)$ (where $m$ is LP1 or LP2),
which obey  driven polaritonic Gross-Pitaevskii equations.
These equations are discussed in the Supplementary Information and  can also be found in the literature, e.g.\ for the single-cavity case in Ref.\ \cite{liew-etal.08} and for the present double-cavity case in Ref.\ \cite{ardizzone-etal.13}.
The interaction between polaritons of the same (opposite) circular polarization, $\tilde{T}^{++}$ ($\tilde{T}^{+-}$),  differs from the excitonic
interactions ${T}^{++}$ and ${T}^{+-}$
 by appropriate combinations of Hopfield coefficients
  because only the excitonic components of the polaritons interact.
   The exact definition for $\tilde{T}^{++}$ and $\tilde{T}^{+-}$,
    as well as that of the
  effective phase-space filling coefficient, $\tilde{A}_{PSF}$, which accounts for polariton interactions due to phase-space blocking of the constituent Fermionic particles (electrons and holes), is given in the
  Supplementary Information.

We include  pump
sources that accounts for the fact that, in the double-cavity, an incident field results in different sources for LP1 and LP2.
We consider only stationary solutions.
After solving the Gross-Pitaevskii equations in configuration space, we perform a Fourier transform and obtain
$\psi^{\pm}_{LP1}(\textbf{k})$, which determines the off-axis emission related to the occupation of the elastic circle,
and define the Stokes parameters, i.e.\ the pseudo-spin, in terms of the LP1 wave functions.
 The $S_1$ component of the pseudo-spin vector is, without conventional normalization denominator,
 $ S_1 ({\bf k} ) =  | \psi^x_{LP1} ({\bf k} )  |^2   -   | \psi^y_{LP1} ({\bf k} )  |^2 $.
  With that normalization included, it reads
 $ S^{(n)}_1 ({\bf k} ) = S_1 ({\bf k} ) / [ | \psi^x_{LP1} ({\bf k} )  |^2   +   | \psi^y_{LP1} ({\bf k} )  |^2 ]$. Alternatively, we can write
 $ S_1 ({\bf k} ) = 2 {\rm Re} [  \psi_{LP1}^{- \ast} ({\bf k})    \psi^{+}_{LP1} ({\bf k}) ]$,
 and furthermore,
 $ S_2 ({\bf k} ) = - 2 {\rm Im} [  \psi_{LP1}^{- \ast} ({\bf k})    \psi^{+}_{LP1} ({\bf k}) ]$
 and
 $ S_3 ({\bf k} ) =  | \psi^+_{LP1} ({\bf k} )  |^2   -   | \psi^-_{LP1} ({\bf k} )  |^2 $.

\begin{figure}
\includegraphics[scale=0.6]{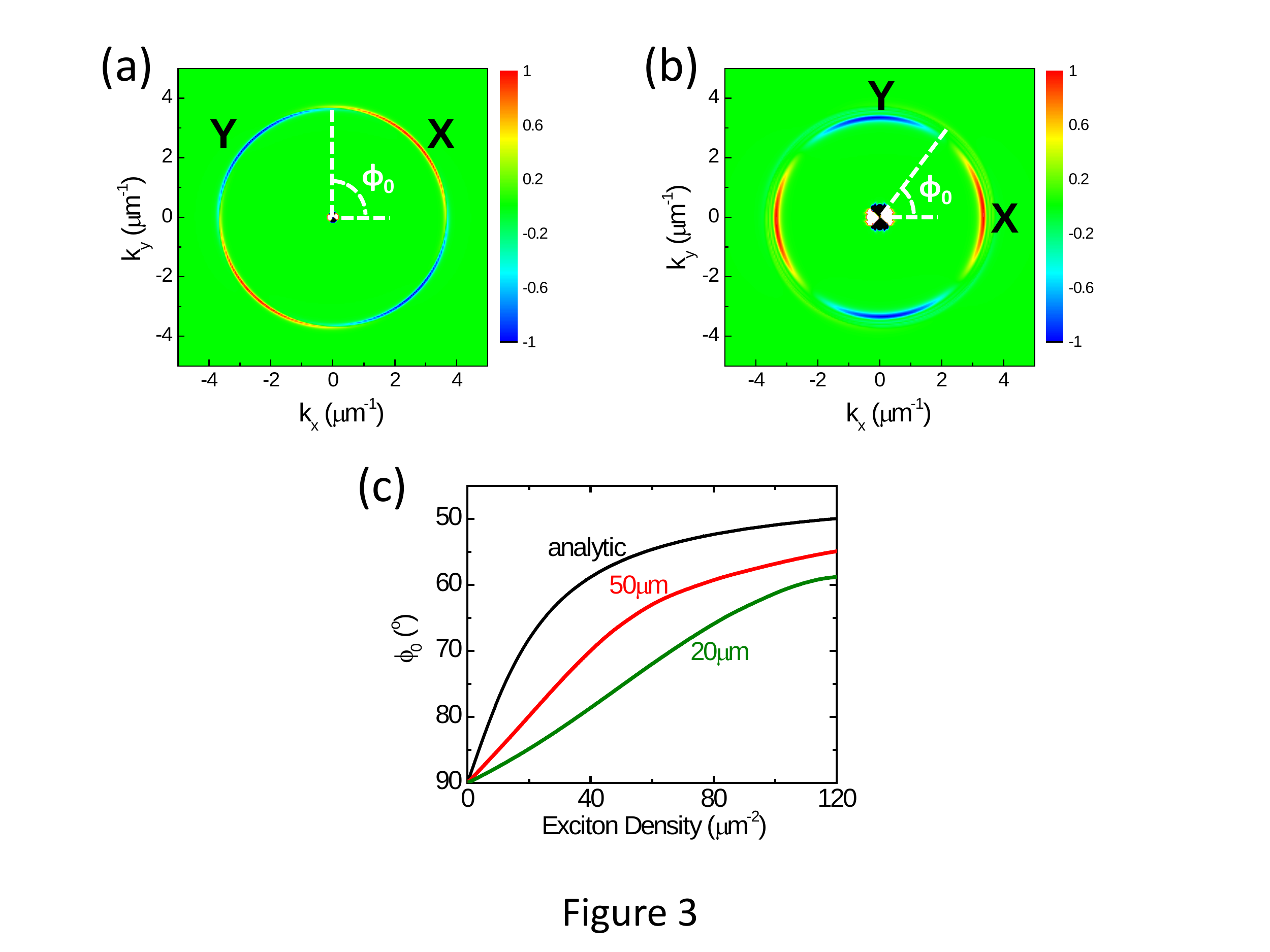}
\caption{
{\bf Theoretical results for the controlled OSHE.}
(a,b) Numerical results for the polarization  $ S_1 ({\bf k})$, with red (blue) indicating  predominantly X (Y)  polarization for a Gaussian pump pulse with diameter 50 $\mu$m (FWHM in intensity).
Here, the data are normalized such that the maximum (minimum) value of $ S_1 ({\bf k})$  on the elastic circles is +1 (-1).
The function  $S_1 ({\bf k})$ contains information about the linear polarization seen in the far field emission.
  Low excitation power shown in (a), high power with an exciton density of $164 \mu$m$^{-2}$ shown in (b).
 The nonlinear OSHE clockwise rotation of about 37$^{\circ}$ is clearly seen.
 (c) Theoretical angular orientation (zero-crossings) $\phi_0$ of the radial average, $ \bar{S}^{(n)}_1$,  from Eq.\ (\ref{equ:angular-shift-simple-model}) and from numerical solutions of the 2-dimensional spinor-polariton Gross-Pitaevskii equation for finite spot diameters $d_{FWHM}$, indicated in the plot.
 The horizontal axis represents the exciton density, which is a fraction of the LP2 polariton density,
 $|p|^2 = \beta_{22} |\psi^{+}_{LP2}(r=0)|^2 $ (see text).
}
\label{fig:k-space-texture}
\end{figure}

Using a low-intensity Gaussian beam profile with FWHM (in intensity) of 50$\mu$m
for the circularly (``+'') polarized source, the steady-state linear polarization texture is shown in
Fig.\ \ref{fig:k-space-texture}a,
exhibiting the linear OSHE effect, in agreement with
 Fig.\ \ref{fig:steady-state-torque}d.
 Since quasi-monochromatic pumping occupies only states on the (hardly distinguishable) elastic circles of the TE and TM modes, the polarization texture is restricted to the vicinity of the circles.
Increasing the pump intensity yields a rotation of the polarization textures as shown in
  Fig.\ \ref{fig:k-space-texture}b,
 again in agreement with the expectations of
Fig.\ \ref{fig:steady-state-torque}d.

 In order to quantitatively determine the rotation of the polarization texture,
we average each term entering the Stokes parameter
along the radial direction over the ring seen in
Fig.\  \ref{fig:k-space-texture}a,b,
and denote the average as
$\langle  | \psi^{x/y}_m ({\bf k} )  |^2  \rangle_{\phi}$, which is a function of the polar angle $\phi$. We can now use these radial averages in the definition of the Stokes parameters, in particular $S_1$, and obtain a Stokes parameter, denoted by $ \bar{S}_1 (\phi)$,
 that depends only on $\phi$.
 $ \bar{S}_1 (\phi)$ (and  the normalized counterpart $ \bar{S}^n_1 (\phi)$)
 displays the expected  $\cos(2 \phi + \Theta)$ dependence, where $\Theta$ is an offset determined the pump field characteristics such
 as pump power. At high pump powers, the far-field pattern rotates with increasing power, i.e. $\Theta$ changes with a power-dependent angular rotation.
We finally denote by $\phi_0$ the angle at which $ \bar{S}_1 (\phi)$ vanishes, i.e.\
$ \bar{S}_1 (\phi_0) \sim \cos(2 \phi_0 + \Theta) =0$. The angle $\phi_0$ is a quantitative measure of the nonlinear OSHE rotation and can be obtained from the numerical solution as well as the experiment, as detailed below. It can also be obtained from analytical models in simple limiting cases. One of those cases is that of a spatially homogeneous excitation (infinite pump beam spot size), which yields
\begin{eqnarray}
 \phi_0 & \approx  & \frac{1}{2} \left(\pi - \arctan   \right. \nonumber  \\
 & & \left.  \left[ ((2 ( \tilde{T}^{++} + \tilde{A}_{PSF} )- \tilde{T}^{+-} ) | \psi^{+}_{LP2} (\textbf{r}=0) |^2 ) / \gamma \right]
 \right) \nonumber \\
&&  \label{equ:angular-shift-simple-model}
\end{eqnarray}
%
This expression allows us to identify the pump-induced $B_3$ component of the effective magnetic field, which was left as a parameter in the  above discussion of the pseudo-spin model, as
$B_3 = (2 (\tilde{T}^{++} + \tilde{A}_{PSF} )- \tilde{T}^{+-} ) | \psi^{+}_{LP2} (\textbf{r}=0) |^2 )$. From a microscopic point of view, the control of $B_3$ is based here on the interaction between LP1 and LP2 polaritons, with the LP2 population giving rise to Coulombic and phase-space filling
 energy shifts of the two LP1 polariton spin states. Since $\tilde{T}^{++}$ and $\tilde{A}_{PSF}$
 are positive and $\tilde{T}^{+-}$ negative, the difference between the shifts due to co-circularly and counter-circularly polarized states leads to a reinforcement, rather than a cancellation, of the nonlinear OSHE rotation.
We also note that the angular rotation in Eq.\ (\ref{equ:angular-shift-simple-model}) changes sign if we use a ``-'' instead of  ``+'' polarized source.

 In Fig.\ \ref{fig:k-space-texture}c
 we show the orientation $ \phi_0 $
 for both the simple
analytical model, Eq.\ (\ref{equ:angular-shift-simple-model}), as well as numerical solutions of the polaritonic Gross-Pitaevskii
equation with finite spot size (details are given in the Supplementary Information).
Since only the excitonic components of the polaritons interact and therefore only these components yield the rotation of the OSHE orientation $\phi_0$,
it is beneficial to show the dependence of $\phi_0$ on
the exciton density per quantum well,
$|p|^2 = \beta_{22} |\psi^{+}_{LP2}(r=0)|^2 $
where the factor $\beta_{22}$ is given in Eq.\ (10) in the Supplemental Information,
rather than the polariton density
$| \psi^{+}_{LP2}  |^2$.
In Fig.\ \ref{fig:k-space-texture}c,
for fixed exciton density, increasing the spot size yields a larger rotation of the OSHE pattern.  We restrict ourselves here to Gaussian pulse profiles and moderate  densities. A detailed analysis of
the OSHE rotation caused by other pump pulse profiles and over a larger range of  densities will be given elsewhere.

\begin{figure}
\includegraphics[scale=0.6]{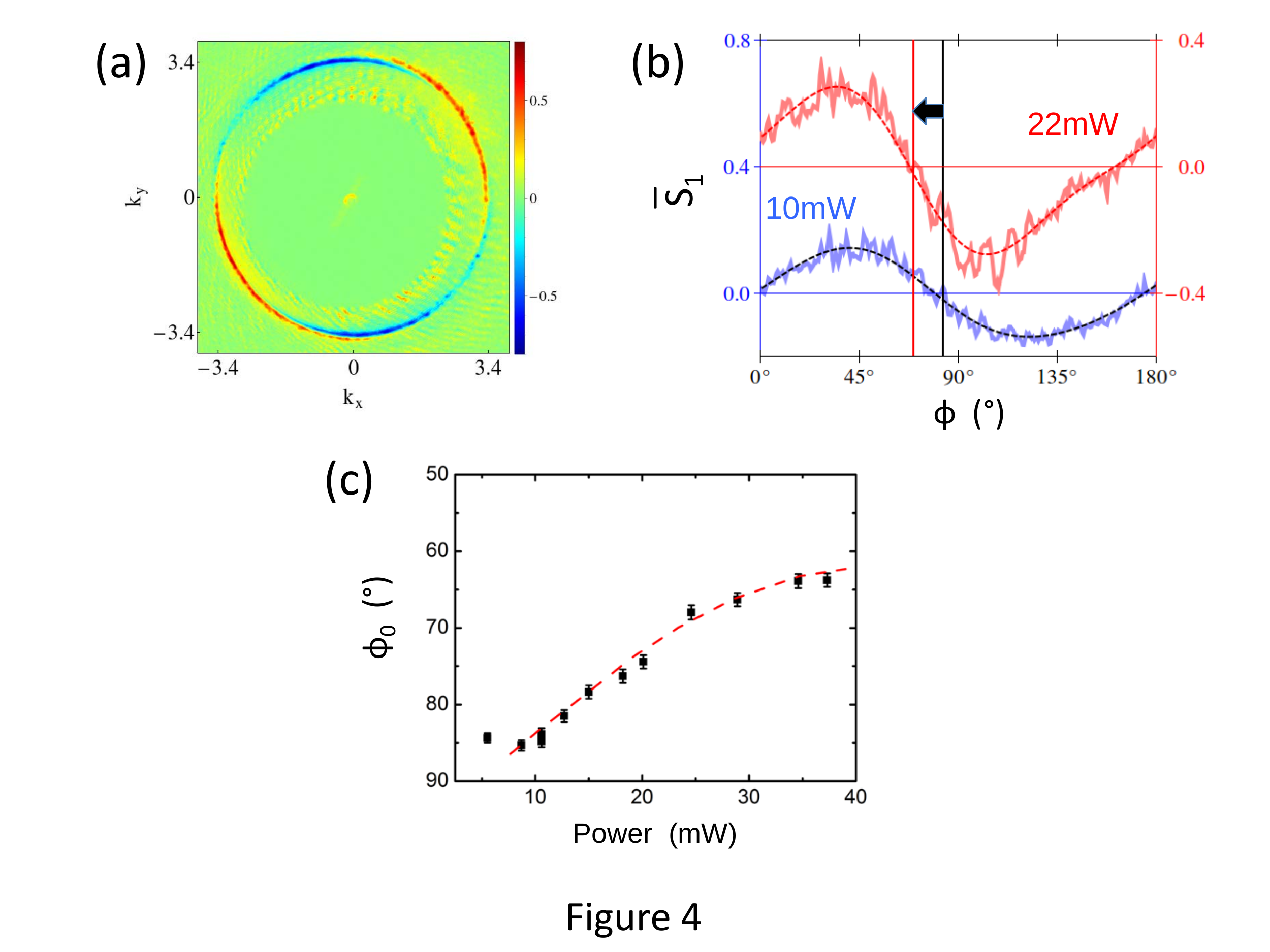}
\caption{
{\bf Experimental results for the controlled OSHE.}
(a) Experimentally observed far-field polarization texture  $ S_1 ({\bf k})$, obtained for a 22mW ``+'' circularly polarized pump incident on the cavity.
(b) Experimental angular variation of the radial average, $ \bar{S}^{(n)}_1$, for two incident excitation powers (blue 10mW, red 22mW) displaying  a light-induced shift of $-13.8^\circ \pm 0.7^\circ$ of the angular orientation as indicated by the positions of the zero crossings (vertical lines). The dashed lines correspond to the fit described in the Data Analysis section.
(c)  Experimental angular orientation (zero-crossings)  $\phi_0$
vs.\ incident excitation power. The thin dashed curve is merely a guide for the eye.
The power range in 4c is estimated to correspond to the range of exciton densities in Fig.\ 3c (50 $\mu$m curve) for
an estimated lifetime of about 100 ps (see text).
}
\label{fig:exp-results}
\end{figure}

Let us now present our experimental results.
 The sample is excited at normal incidence with a circularly polarized mono-mode laser.  An example of the observed far-field pattern $ S^{(n)}_1 ({\bf k} )$ is shown in
 Fig.\ \ref{fig:exp-results}a.
 The corresponding radial average shows the predicted $\cos(2 \phi + \Theta)$ dependence, as shown in
Fig.\ \ref{fig:exp-results}b
for two different powers. We see that the curves shift with increasing intensity, for example by comparing the zero crossings of the two curves $\phi_0$ (indicated by vertical lines).
 In
 Fig.\ \ref{fig:exp-results}c
 we plot these positions as a function of power. The experimental data confirm the theoretical predictions and agree very well with Fig.\ 
  \ref{fig:k-space-texture}c.
 We have verified that the power range shown in
  Fig.\ \ref{fig:exp-results}c

 corresponds approximately to that shown in
 Fig.\ \ref{fig:k-space-texture}c
 for the 50$\mu$m results.   An equation for energy conservation of the cavity, balancing incident, reflected and transmitted powers with loss through exciton recombination, can be cast in the form
 $P_{inc} \sim \gamma_{rec}   |p|^2 $
 where $P_{inc}$ is the incident power, $ |p|^2 $ the exciton density in each quantum well (approximately the same for all quantum wells), $\gamma_{rec}$ the recombination rate (including radiative and non-radiative recombination),
 and the proportionality constant (omitted here) is given in Eq.\ (24) of the Supplemental Information.
 The scales agree if the lifetime $\tau_{rec} = \hbar / \gamma_{rec}$ is approximately 100ps, which, as discussed in the
   Supplemental Information, agrees with the value expected for our cavity.

 %
 %
 %

We finally note that
we have verified that the experiment displays no bistable behaviour, that, as predicted by theory,
the rotation  is reversed when the pump's circular polarization is reversed, and that
 the effect is also observed with the same rotation angles in the diagonal polarization basis. \\

{\centerline  {\bf Discussion}}

In conclusion,
the experimental results, taken together with the theoretical analysis, give a clear indication of the direct control of the optical spin Hall effect.
 The good agreement between the experimental and theoretical results for the far field proves that the local polarization texture of the polaritonic quantum fluid has been modified all optically.
 %
%
This reinforces the widely held belief that
the light-matter nature of exciton-polaritons is ideal for the development of spinoptronic devices, in which the control of the optical excitation is at the heart of the device operation. It also lays out a path towards relatively simple and robust designs as an alternative to the polaritonic Bose condensates, in which a plethora of polaritonic interaction effects and their control is already well-established.
Extensions of our work  including multiple beams and/or sequences of short pulses may in the future allow for an on-demand generation of a large selection of spin textures and currents. \\

{\centerline {\bf Methods}}

{\bf Sample design and experimental setup}

The sample consists of two coupled $\lambda/2$ Ga$_{0.05}$Al$_{0.95}$As cavities embedded between three Ga$_{0.05}$Al$_{0.95}$As/Ga$_{0.8}$Al$_{0.2}$As Bragg mirrors with 25 (back), 17.5 (middle), and 17.5 (front) pairs respectively.
The nominal Q factor is around $10^5$ and the middle Bragg mirror induces an approximately 10meV coupling between bare cavity modes. In each cavity 3 sets of four 7 nm GaAs QWs are inserted at the antinodes of the  field resulting in a 13 meV Rabi splitting.
During growth of the structure by molecular beam epitaxy, a wedge is introduced in the cavities thickness allowing to tune the cavity modes energies with respect to the bare excitonic mode energy.

Experiments are performed at temperature 6 K, the excitation laser is provided by a single mode Ti:Sapphire laser.
The polarization-resolved far-field spectroscopy is obtained by projecting the Fourier plane of the detection objective on a cooled CCD camera.
The pump beam is hidden by a spatial filter.

{\bf Data analysis}

Prior to the analysis, light observed on both linear polarization channel $X$ and $Y$ on the elastic circle is integrated over momentum amplitude (range 3.16 - 3.64 $\mu$m$^{-1}$).
The degree of linear polarization is obtained by
$\bar{S}^{(n)}_1=\frac{X-Y}{X+Y}$
for the obtained intensity vs angle curves. 
Data are fitted using a multipolar development up to eighth order : $A_d \cos ( \phi + \theta_D) + A_q \cos (2 \phi + \theta_Q) + A_o \cos (4 \phi + \theta_O)$ where sinusoidal terms are dipolar, quadrupolar and octopolar terms in consecutive order. 
At the lowest order in the effective magnetic field perturbation, only  the quadrupolar  phase is modified, while the various sources of experimental imperfections (excitation and detection polarization miscalibration, built-in stress, optical misalignment, $k=0$ position on image) do not affect the quadrupolar phase except for spherical aberrations of the imaging setup and polarization-dependent elastic scatterings, which are intensity-independent.
The data are slightly heteroscedatic since photon noise is proportional to the average intensity,
but heteroscedasticity-consistent standard errors\cite{White1980} are similar to standard errors from a simple linear fit using a linearized model. The standard errors obtained from fitting are multiplied by a factor $\sqrt{\beta_c/\beta_d}$ where $\beta_c=3.7^\circ$ is the characteristic autocorrelation angle of residuals, and $\beta_d=0.6^\circ$ is the discretization step angle of data to account for spatial correlation effects on fitting results.
The 68\% expanded uncertainty indicated by the error bars in Fig.\ 4c takes into account the experimental accuracy of 0.6$^{\circ}$ due to waveplate positioning. \\




{\centerline  {\bf Addendum}}

 We gratefully acknowledge financial support from the US NSF under grant ECCS-1406673, TRIF Photonics, and the German DFG (research centre TRR142 ‘Tailored nonlinear photonics’ and SCHU 1980/5-1 and 5-2). S.Sch. further acknowledges
support through the DFG Heisenberg programme. We thank Kyle Gag for helpful discussions.

 The authors declare that they have no
competing financial interests.

 Correspondence and requests for materials
should be addressed to R.B.\  \\ (email: binder@optics.arizona.edu).

\end{document}